\documentclass[journal]{IEEEtran}
\usepackage{graphicx}
\usepackage{subfigure}
\usepackage{float}
\usepackage{subfloat}
\usepackage{amsmath}
\usepackage{amssymb}
\usepackage{amsthm}
\usepackage{epstopdf}
\usepackage{cases}
\usepackage{color}
\usepackage{makecell}
\usepackage{cite}
\usepackage{algorithmic}
\usepackage[ruled]{algorithm2e}

\usepackage[normalem]{ulem}

\makeatletter
\renewcommand\normalsize{%
	\@setfontsize\normalsize\@xpt\@xiipt
	\abovedisplayskip 6.5\p@ \@plus2\p@ \@minus5\p@
	\abovedisplayshortskip \z@ \@plus3\p@
	\belowdisplayshortskip 6.5\p@ \@plus3\p@ \@minus3\p@
	\belowdisplayskip \abovedisplayskip
	\let\@listi\@listI}
\makeatother

\ifCLASSINFOpdf
\else
\fi
\usepackage{caption}
\usepackage{mathtools}

\begin{document}

%
\title{Integrated Communication, Sensing, and Computation Framework for 6G Networks}
%
%
%

\author{Xu Chen,~\IEEEmembership{Member,~IEEE,}
	   Zhiyong Feng,~\IEEEmembership{Senior Member,~IEEE,}
	   J. Andrew Zhang,~\IEEEmembership{Senior Member,~IEEE,}
	   Zhaohui Yang,~\IEEEmembership{Member,~IEEE,}
	   Xin Yuan,~\IEEEmembership{Member,~IEEE,}
	   Xinxin He,~\IEEEmembership{Member,~IEEE,}
	   and Ping Zhang,~\IEEEmembership{Fellow,~IEEE}
	   
	   \thanks{X. Chen, Z. Feng, X. He are with School of Information and Communication Engineering, Beijing University of Posts and Telecommunications, Beijing 100876, P. R. China (Email:\{chenxu96330, fengzy, hxx\_9000\}@bupt.edu.cn).}
	   
	   \thanks{J. A. Zhang is with the Global Big Data Technologies Centre, University of Technology Sydney, Sydney, NSW, Australia (Email: Andrew.Zhang@uts.edu.au).}
    
\thanks{Z. Yang is with the College of Information Science and Electronic Engineering, Zhejiang University, Hangzhou 310007, China, and Zhejiang Provincial Key Lab of Information Processing, Communication and Networking (IPCAN), Hangzhou 310007, P. R. China (Email: yang\_zhaohui@zju.edu.cn).}
	   
	   \thanks{X. Yuan is with the Commonwealth Scientific and Industrial Research Organization (CSIRO), Australia (Email: Xin.Yuan@data61.csiro.au).}
	   
	   \thanks{Ping Zhang is with Beijing University of Posts and Telecommunications, State Key Laboratory of Networking and Switching Technology, Beijing 100876, P. R. China (Email: pzhang@bupt.edu.cn).}
	   \thanks{Correponding author: {Zhiyong Feng}}
	   }


%
%

\markboth{}%
{Shell \MakeLowercase{\textit{et al.}}: Bare Demo of IEEEtran.cls for IEEE Journals}
%



\maketitle

\pagestyle{empty}  
\thispagestyle{empty} 

\begin{abstract}
In the sixth generation (6G) era, intelligent machine network (IMN) applications, such as intelligent transportation, require collaborative machines with communication, sensing, and computation (CSC) capabilities. This article proposes an integrated communication, sensing, and computation (ICSAC) framework for 6G to achieve the reciprocity among CSC functions to enhance the reliability and latency of communication, accuracy and timeliness of sensing information acquisition, and privacy and security of computing to realize the IMN applications. Specifically, the sensing and communication functions can merge into unified platforms using the same transmit signals, and the acquired real-time sensing information can be exploited as prior information for intelligent algorithms to enhance the performance of communication networks. This is called the computing-empowered integrated sensing and communications (ISAC) reciprocity. Such reciprocity can further improve the performance of distributed computation with the assistance of networked sensing capability, which is named the sensing-empowered integrated communications and computation (ICAC) reciprocity. The above ISAC and ICAC reciprocities can enhance each other iteratively and finally lead to the ICSAC reciprocity. To achieve these reciprocities, we explore the potential enabling technologies for the ICSAC framework. Finally, we present the evaluation results of crucial enabling technologies to show the feasibility of the ICSAC framework.

\end{abstract}


%
\IEEEpeerreviewmaketitle


\begin{figure*}[!t]
	\centering
	\includegraphics[width=0.45\textheight]{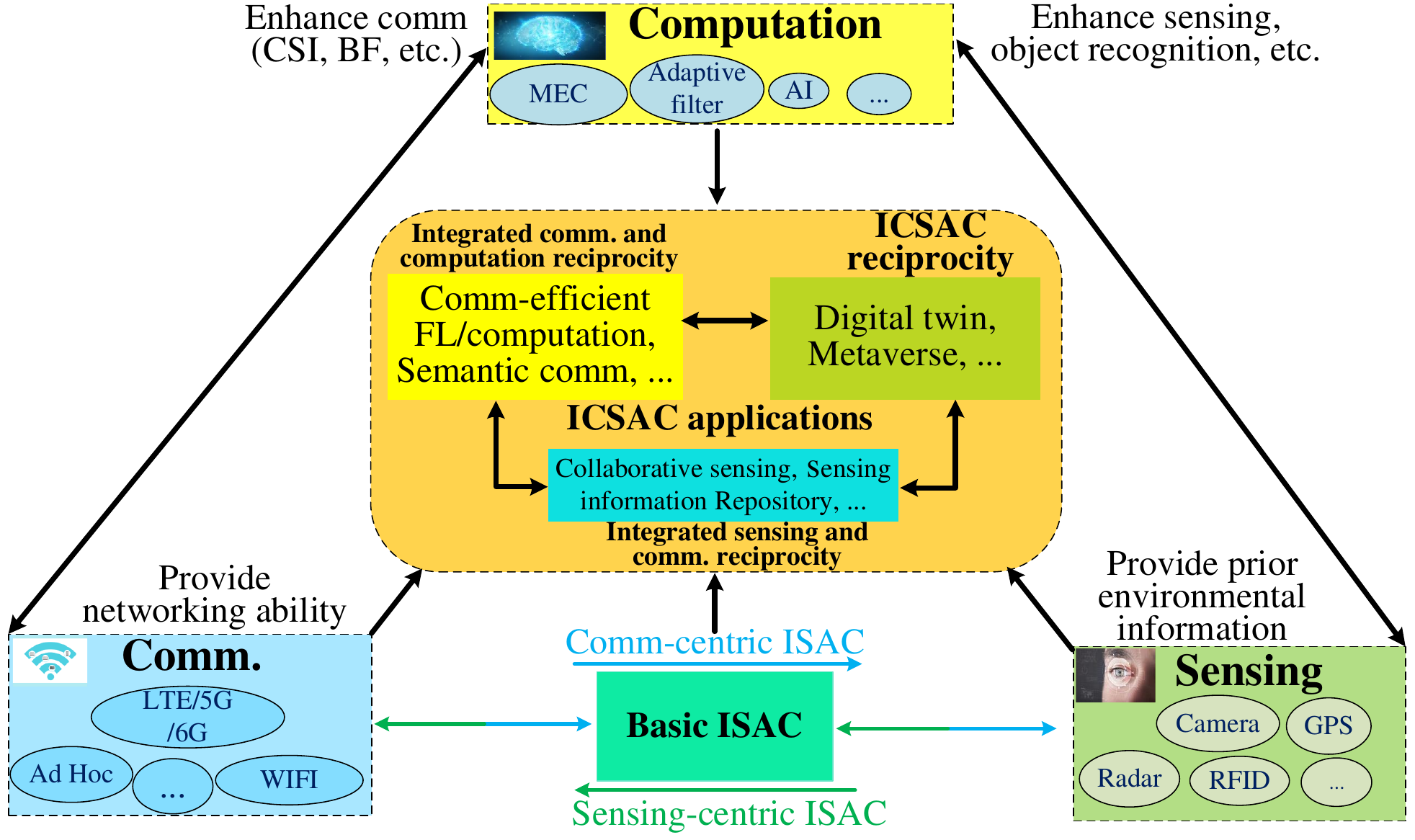}%
	\DeclareGraphicsExtensions.
	\caption{ICSAC framework comprising communication, sensing, and computation functions, where the sensing function perceives the environmental information to provide prior knowledge for communication optimization and application decision-making; the computation function processes the sensing information to generate instructions for controlling actuators; and the communication function links different functions, transmitting data to processors and instructions to actuators.}
	\label{fig:JCSC_framework}
\end{figure*}

\section{Introduction}

In the future sixth generation (6G) era, intelligent machine network (IMN) applications, such as intelligent vehicular networks, smart cities, and intelligent manufacturing, require intelligent machines (IMs) to possess autonomous environmental perception and collaboration abilities to perform complex tasks such as motion control~\cite{You2021}. To achieve intelligent applications, both communication and sensing capabilities are essential. Moreover, the cooperation among IMs will also enable multiple computation technologies, such as collaborative computing~\cite{Zhao2023}. Integrated sensing and communications (ISAC) technology has already emerged as a promising technique and has been recognized as one of the six key scenarios and technologies for 6G by IMT-2030~\cite{IMT2030_Recommendations}.

Recently, there have been new trends in wireless networks~\cite{liu2020joint}. For example, there have been studies in exploiting the intelligent computing capability to achieve joint gains between communication and sensing, thereby improving sensing accuracy and communication reliability~\cite{Chen2023Kalman}. Moreover, utilizing networking capabilities to achieve distributed intelligent signal processing, such as federated learning (FL), has also been a promising method to improve computing efficiency and the security of user privacy. These trends indicate the potential for mutual gains among communication, sensing, and computation (CSC) functions~\cite{Feng2021Nov}. We refer to the signal processing techniques and networking protocols that achieve joint gains among CSC as integrated communication, sensing, and computation (ICSAC) technology. 

To support future wireless communications, there are several challenges in achieving 6G IMN applications, which are summarized as follows:
\begin{itemize} 
	
	\item The 6G applications demand extremely high requirements for communication reliability (10$^{-5}$ to 10$^{-7}$), latency (sub-ms level), and sensing accuracy (sub-decimeter level)~\cite{IMT2030_Recommendations}.
	
	\item The scenarios and tasks of 6G are diverse and highly dynamic, especially in IMN scenarios. The dynamic features include highly dynamic channel state information (CSI) due to high mobility and rapidly changing network performance requirements due to time-varying task states or application demands~\cite{Feng2021Nov}.
	
	\item 6G IMN applications require closed-loop control, which involves a complete application control process, e.g., environmental sensing, sensing information transmission, decision command dissemination, and environmental sensing feedback~\cite{Feng2021Nov}. The entire closed-loop control process requires high reliability and low latency for each step while minimizing the task completion time.
	
\end{itemize} \par

To solve the above three challenges of 6G networks, and meet the aforementioned trends in the ICSAC technology study, this article proposes the ICSAC framework to achieve joint gains in communication, sensing, and computation. The main innovations of this framework are summarized as follows:

\begin{itemize} 
	
	\item We conduct an in-depth analysis of CSC functions, elucidating the mechanism of mutual gains in CSC capabilities. This enables the proposed ICSAC framework to be a guiding map for innovations in wireless signal processing. By exploiting the rapidly evolving intelligent computing capabilities to excavate the sensing information with high timeliness as prior information, we can enhance communication reliability and network routing performance, as well as reduce latency. The enhanced communication networking capabilities further improve the performance of distributed computing architectures. The improvement in computing and communication networking capabilities, in turn, can further enhance the performance of networked sensing. 
	
	\item The joint gains of communication, sensing, and computation in the ICSAC framework are beneficial for meeting the challenge of highly-dynamic task requirements of 6G IMN. In particular, ISAC enables simultaneous wireless communication and environmental sensing. The ICSAC network can leverage the sensing information with high timeliness to perform rapid intelligent network optimization using artificial intelligence (AI) technologies such as deep learning (DL) and reinforcement learning (RL). This enables the network to adapt to the rapid changes of the environment and requirements of 6G network scenarios.
	
	\item The joint gains of CSC in the ICSAC framework can be used in closed-loop control applications. These applications require closed-loop optimization of the sensing information acquisition and transmission, decision-making based on sensing information, decision command dissemination, and environmental sensing feedback. The proper design of CSC functions in the ICSAC framework can comprehensively improve sensing accuracy, communication reliability and latency, and computation efficiency. This enhances the network's robustness, reduces data retransmission, and decreases the latency in each stage of the closed-loop control application.
	
\end{itemize} \par

The ICSAC technology can achieve mutual benefits in CSC functions, enhancing the efficiency and reliability of communication, networking, sensing, and computation. Therefore, it is promising to become one of the mainstream orientations in 6G areas. 

The remaining parts of this article are arranged as follows. Section \ref{sec:framework} introduces the ICSAC framework and its mechanism of achieving joint communication, sensing, and computation gains. Section \ref{sec:key_technology} elaborates on the key enabling technologies and challenges of the ICSAC framework. Section \ref{sec:evaluation} presents the evaluation performance of several key technologies to show the feasibility of the ICSAC framework. Section \ref{sec:conclusion} concludes the entire article and provides an outlook on possible future work.

\begin{figure*}[!t]
	\centering
	\includegraphics[width=0.45\textheight]{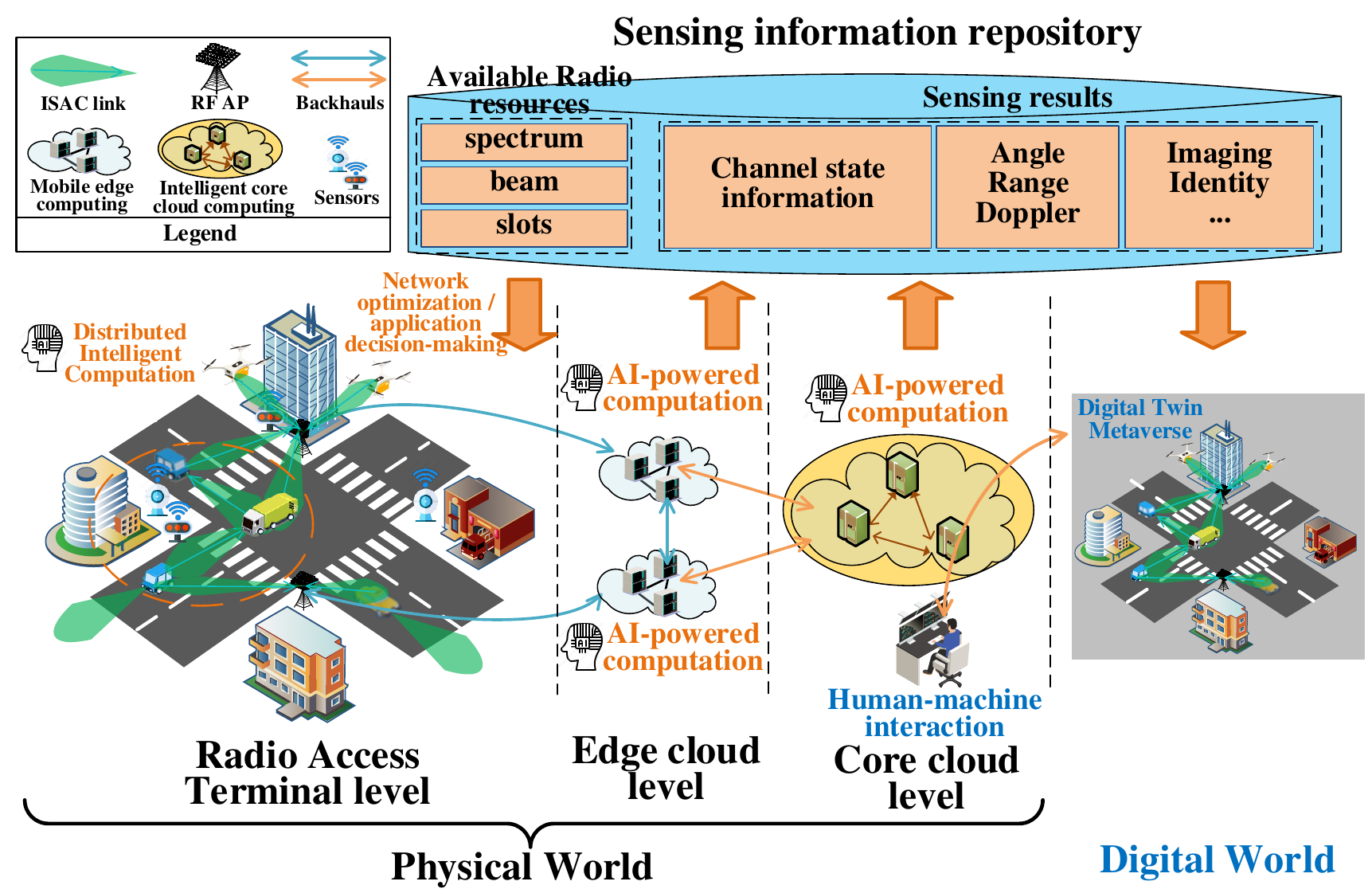}%
	\DeclareGraphicsExtensions.
	\caption{An illustration of the ICSAC framework.}
	\label{fig:JCSC_scenario_framework}
\end{figure*}

\section{ICSAC Framework}\label{sec:framework}

To achieve mutual enhancement among CSC functions and satisfy the requirements of 6G networks for high-reliability and low-latency communication, and high-accuracy sensing, we propose an ICSAC framework, as shown in Fig.~\ref{fig:JCSC_framework}.

The CSC functions are analog to human sensing organs, neural systems, and the brain, respectively. The sensing function perceives the environmental information to provide prior knowledge for decision-making. Then, the computation function processes environmental information and generates instruction signals based on application intents to control actuators to interact with the environment. To build the connection between the sensing and computation functions, the communication function transmits the environmental information to computation processors and instructions to the actuators.

Due to the complex intracell and intercell interference in wireless networks, it is necessary to optimize wireless resource allocation and use anti-noise signal processing methods to improve the quality of sensing information acquisition (such as sensing resolution and accuracy) as well as the reliability and capacity of communication. The intelligent computing power enabled by AI technologies, such as DL and RL, will significantly enhance the performance of sensing processing and improve the communication optimization with the assistance of the sensing prior information~\cite{Zhang20196G}, which is called the computing-empowered ISAC reciprocity. In addition, based on reliable communication links and sensing information, nodes can form networks for distributed computation and cooperative sensing, which is called the sensing-empowered integrated communication and computation (ICAC) reciprocity. The above ISAC and ICAC reciprocities can enhance each other iteratively, leading to the combined ICSAC reciprocity.

Then, we introduce the main techniques of CSC functions and elaborate on the reciprocity among the CSC functions of the ICSAC framework.

\begin{figure*}[!t]
	\centering
	\includegraphics[width=0.6\textwidth]{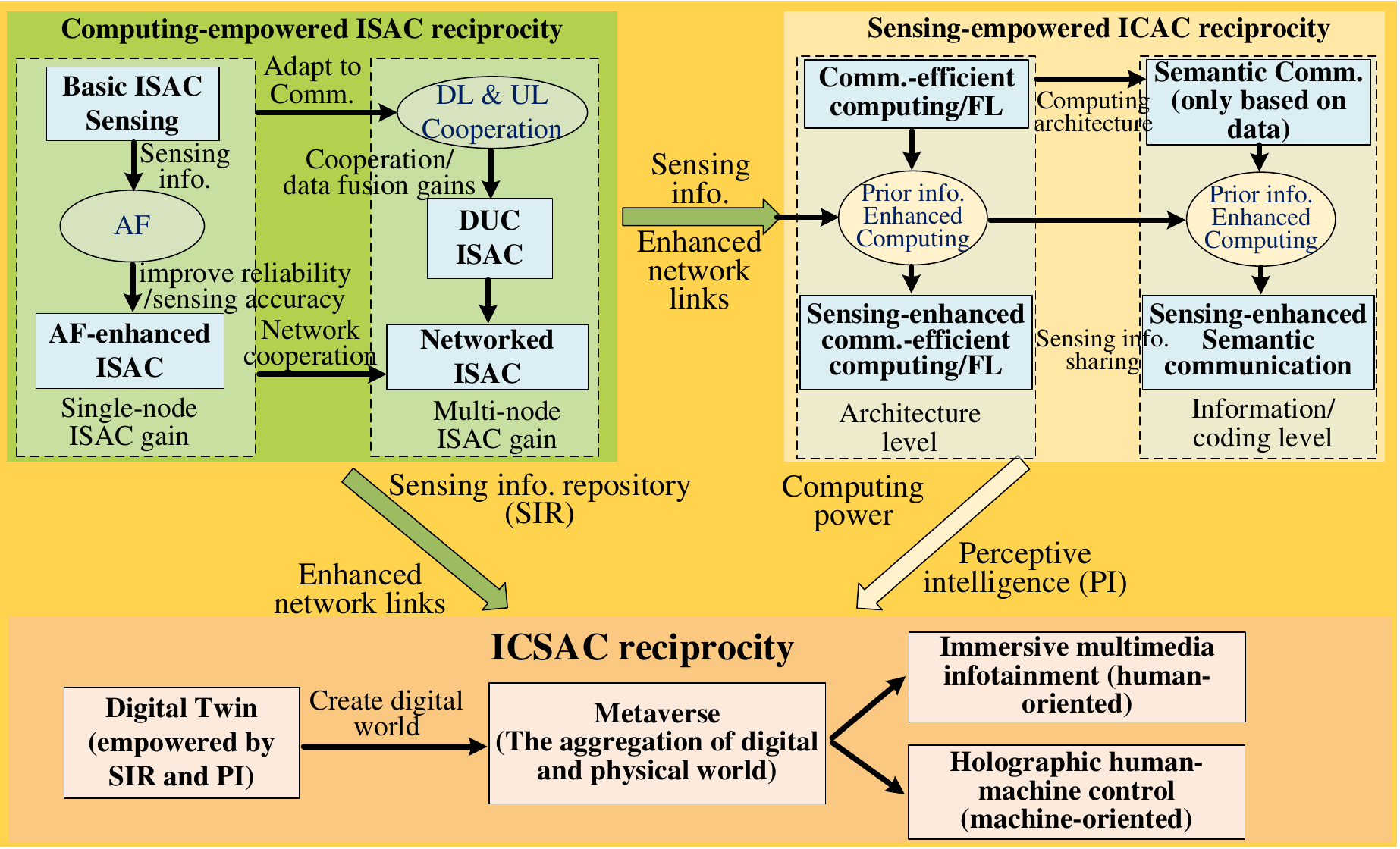}%
	\DeclareGraphicsExtensions.
	\caption{The illustration of the potential enabling technologies for the ICSAC framework.}
	\label{fig:ICSAC_framework_illustration}
\end{figure*}

\subsection{Main Techniques of Communication, Sensing, and
Computation}

\textbf{Sensing}: We consider the sensing function of machines as ubiquitous spectrum sensing. For example, camera sensors capture physical images from visible-light spectrum information. Besides, radar sensing can acquire physical information, such as angles, distances, and speeds, based on radio spectrum sensing. Furthermore, GPS can acquire location information. The above sensing information can be collected to form a sensing information repository that provides crucial prior environment knowledge for decision-making and network optimization.

\textbf{Communication}: Wireless communication has various regimes, among which mobile cellular communication networks and WiFi are the most dominant and both adopt the multi-input multi-output orthogonal frequency division multiplex (MIMO-OFDM) transceiver architecture. Wireless communication provides networking capabilities for multiple nodes to collaboratively form distributed systems.

\textbf{Computation}: The computation function is generally diverse. Centralized and distributed computation architectures coexist in a network system. The distributed computing architectures enable local decision-making and edge network optimization. Additionally, AI techniques can provide intelligent processing tools for decision-making, network optimization, and signal processing~\cite{DL4CEandBF2022} empowered by prior sensing information.

\subsection{ICSAC Framework} \label{sec:ICSAC_reciprocity}
We use a typical application scenario of intelligent ground-air integrated vehicular networks (GAI-VeNet) to illustrate the reciprocities of communication, sensing, and computation in the ICSAC framework, as shown in Fig. 2. Note that the vehicles refer to both the unmanned aerial vehicles (UAVs) and unmanned terrestrial vehicles (UTVs). Without loss of generality, we refer to the vehicles as machine users (MUEs). The ICSAC framework contains two worlds, i.e., the physical world containing all the entities and environment of the networks, and the digital world which is the digital twin of the physical world. The physical world is composed of three levels, i.e., the radio access terminal (RAT) level, edge cloud level, and core cloud level.

RAT level contains all the MUEs and access points (APs) in the physical world. Wireless users and APs can utilize ISAC techniques to achieve communication and sensing simultaneously. The wireless sensors also perceive the network environment. The neighbored MUEs or APs can form distributed intelligent computing networks to preprocess the sensing information efficiently. APs can collect the sensing information of MUEs and sensors, upload it to the edge cloud servers, or propagate the action instructions from higher levels to the RAT level.

The edge and core cloud levels are composed of the edge and core cloud server networks, respectively. The edge cloud level controls the operation of local wireless networks and the decision-making for concrete applications such as trajectory planning for the local MUEs. Moreover, the edge cloud level can excavate the sensing data collected from the RAT level and form local sensing information repositories providing prior information for network control and decision-making. The core cloud level converges all the sensing information repositories of edge servers to construct a global one that provides prior information for global network control and application decision-making. Moreover, the core cloud level uses AI techniques to form the digital twin of the physical world based on the global sensing information repository and provides the interaction interface to human users (HUEs) for human-machine interaction over the digital metaverse.

\subsection{Reciprocity among the communication, sensing, and computation capabilities} \label{sec:Reciprocity_CSC}

We proceed to elaborate on the reciprocity among the communication, sensing, and computation achieved by the above ICSAC framework. 

\subsubsection{Computing-empowered ISAC Reciprocity} 
The convergence of radio communication and sensing in terms of spectrum, transceivers, and digital signal processing lays the foundation for the hardware implementation of ISAC. ISAC achieves single-base active sensing by processing the autocorrelation between communication signal echoes and transmitted signals or achieves bi-static sensing by processing the received communication signals~\cite{Chen2021CDOFDM, Ni2021ULSen}. 

The communication networking capability can be utilized to realize networked sensing for ISAC nodes or sensors, which can efficiently solve insufficient sensing problems, such as obstruction in the single-node sensing scenario. Due to the diversity of the resources of sensing information, the obtained sensing information exhibits significant multi-modal characteristics. Specifically, the angles, locations, velocities of MUEs and environmental scatterers, CSI estimation, images, etc., are aggregated via the communication network at both the edge and core cloud levels, forming local and global sensing information repositories, respectively. CSIs, user locations, velocities, and other related information are useful for communication network optimization and application decision-making. 

\subsubsection{Sensing-empowered ICAC Reciprocity}
By exploiting communication networking capabilities, multiple nodes can collaborate to train distributed neural networks or other distributed computing architectures. 
Distributed learning enhances the efficiency and scalability of the training process, which is especially useful when dealing with massive datasets that would be impractical to transfer and process centrally.
Since each network node is trained using its local data and only transmits model parameters without exposing the private data, distributed computing architectures can protect user data privacy and security. 

There can be parallel multi-tasking requirements in the local computing network, such as the simultaneous transmission of sensing data and the relay of user communication data. In such cases, distributed intelligent computing can intelligently assign priorities of different tasks to each computing node based on the CSI estimation and the locations of information source and destination contained in the sensing information repository, thereby enhancing the efficiency of parallel task completion. Moreover, by using the semantic characteristics hidden within CSIs, transmitted signals, and sensing results, the intentions of machine applications can be disseminated more efficiently among MUEs, omitting the process of extracting human-type semantics based on sensing data for application control.

\subsubsection{ICSAC Reciprocity}
Intelligent computing technologies such as neural networks can be exploited to enhance the anti-noise performance of communication and sensing, improving CSI estimation accuracy, communication reliability, and sensing accuracy. By utilizing the enhanced MUEs’ location and CSI information as prior information, intelligent computing methods such as DL and RL can be used to improve the performance of network resource allocation, routing, beamforming, etc. The improvement in communication performance will further provide better reliability, capacity, and timeliness for networked sensing and distributed intelligent computation, thus achieving iterative ICSAC gain, which is beneficial for closed-loop applications that require low latency and high reliability.

By exploiting AI techniques to digest and enhance the multi-source sensing data, a high-quality digital twin can be achieved based on the global sensing information repository, which lays the foundation for realizing a highly immersive metaverse for HUEs. 

\section{Key Enabling Technologies and Main Challenges} \label{sec:key_technology}

This section explores the key technologies and the corresponding main challenges for realizing the aforementioned reciprocity of the ICSAC framework, including the computing-empowered ISAC, sensing-empowered ICAC, and ICSAC techniques, which are concluded in Fig.~\ref{fig:ICSAC_framework_illustration}. The computing-empowered ISAC techniques offer sensing information and enhanced network links for realizing sensing-empowered ICAC and ICSAC techniques, while the sensing-empowered ICAC techniques provide computing architectures and perceptive intelligence for achieving the ICSAC techniques.

\subsection{Computing-empowered ISAC Techniques} \label{sec:ISAC_techniques}

This subsection presents the crucial techniques for achieving computing-empowered ISAC reciprocity in Section~\ref{sec:ICSAC_reciprocity}. We first introduce the basic ISAC sensing scheme and then elaborate on the adaptive filter (AF)-enhanced CSI enhancer for ISAC, downlink uplink cooperative (DUC) ISAC, and computing-empowered networked sensing.

\subsubsection{Basic ISAC Sensing Scheme}
Since both 6G and WIFI adopt the MIMO-OFDM-based regimes in the future, we focus on the MIMO-OFDM-based ISAC sensing scheme. The most prevalent ISAC sensing scheme is the fast Fourier transform (FFT)-based sensing scheme that applies 2D-FFT to obtain the sensing spectrum for estimating range and Doppler. This method has the lowest complexity, but its on-grid detection nature restricts its sensing accuracy by the range and Doppler resolutions defined by the number of subcarriers and OFDM symbols, respectively~\cite{Chen2023DLISAC}. Besides, zero-padding FFT could alleviate this problem but introduce a large complexity increase. The subspace-based ISAC sensing scheme based on the super-resolution sensing method, such as multiple signal classification (MUSIC), can improve the sensing accuracy but with relatively high complexity when jointly estimating 2D range and Doppler~\cite{Chen2023DLISAC}. 


We can choose the appropriate sensing scheme for ISAC by comprehensively considering the requirements of the applications on sensing accuracy and complexity.

\subsubsection{AF-enhanced ISAC}

AF, such as the Kalman filter, Bayesian filter, and deep learning-based filter, can be used to enhance the CSI estimation for improving communication reliability by exploiting the angle-of-arrival (AoA), range, and Doppler estimation as the prior information, as well as suppressing the timing offset (TO) in the CSI to improve sensing accuracy by exploiting the Doppler estimation with carrier frequency offset (CFO) as the prior information~\cite{CX_ULISAC_2022}. This method can improve both communication and sensing performance significantly with the feature of linearity and is thus promising to be implemented in the ICSAC framework for 6G. 


\subsubsection{DUC ISAC}

In mobile communication networks, the uplink (UL) and downlink (DL) channels between AP and MUE have time-division duplex (TDD) reciprocity. Within adjacent uplink and downlink time slots, the channel environment between the AP and UE remains unchanged. The active sensing conducted by AP in the downlink time slots and the bi-static sensing in the uplink time slots are independent sensing processes of the same physical environment. Therefore, we can improve the sensing performance via data fusion methods and enhance communication performance by exploiting the sensing information obtained in uplink and downlink ISAC as prior information~\cite{Chen_DUC_ISAC}.


\subsubsection{Computing-empowered Networked ISAC}

By exploiting networked collaboration capabilities, multiple distributed sensors (including MUEs and APs that use ISAC) can cooperatively sense the same physical area, effectively reducing blind spots caused by obstruction and random deep fading in the single-sensor scenario~\cite{Jiang2022multiple}. Using distributed computing such as FL, collaborative MUEs can conduct distributed training, where they only need to exchange model parameters to share semantic information, thereby improving the efficiency of the sensing network and ensuring the security of data.

\textbf{Challenges to the Computing-empowered ISAC techniques:} 6G networks may encounter extremely dynamic scenarios characterized by rapidly changing environments. Therefore, all the above techniques necessitate the acquisition of sensing information with exceptional timeliness. Moreover, effective interference management and synchronization for networked sensing are also great challenges.

\begin{figure*}[!t]
	\centering
	\includegraphics[width=0.8\textwidth]{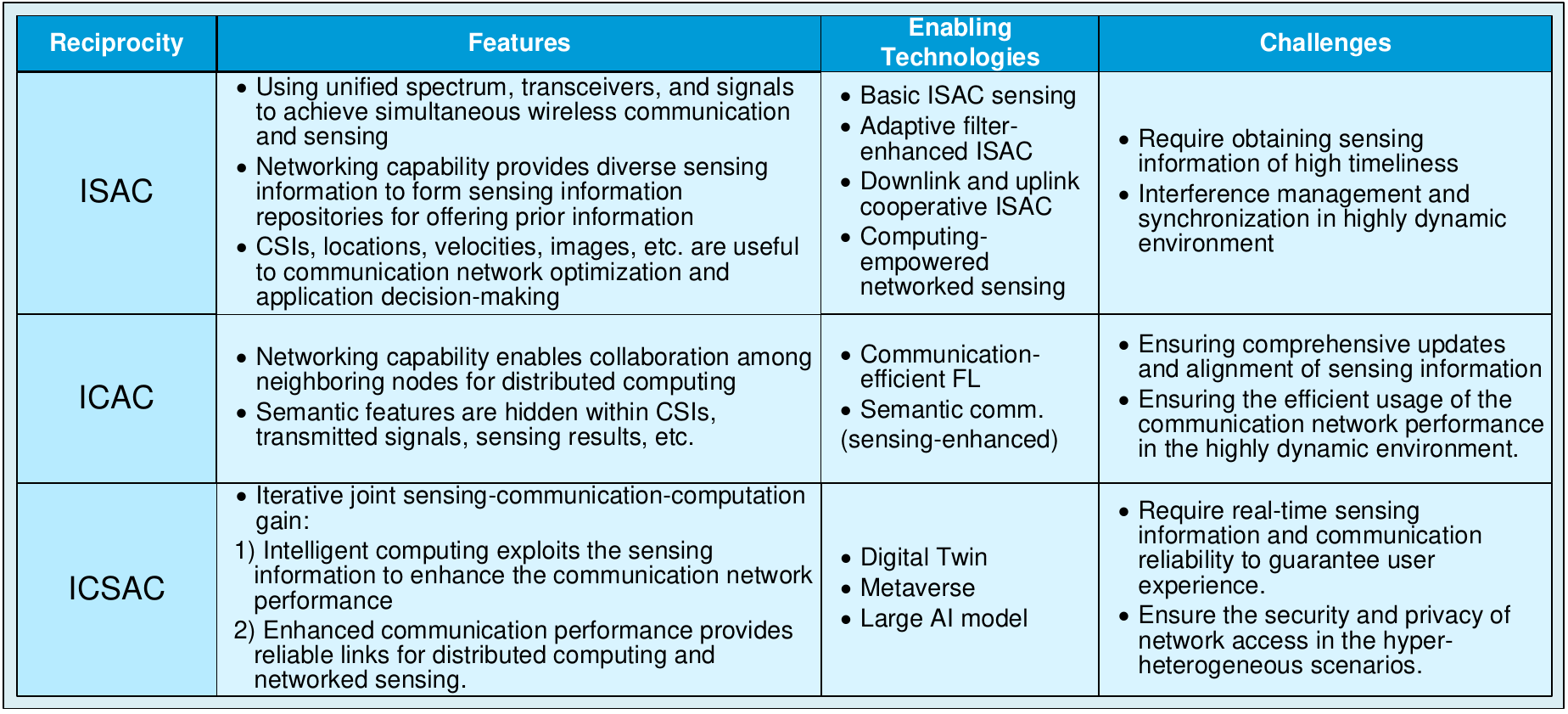}%
	\DeclareGraphicsExtensions.
	\caption{The key enabling technologies and challenges corresponding to the reciprocity of the ICSAC framework.}
	\label{fig:ICSAC_table}
\end{figure*}

\subsection{Sensing-empowered ICAC Techniques} \label{sec:ICAC_techniques}

This subsection presents the critical techniques for realizing the ICAC reciprocity in Section~\ref{sec:ICSAC_reciprocity}, including the sensing-enhanced communication-efficient FL and semantic communication techniques, which can achieve perceptive intelligence.

\subsubsection{Sensing-enhanced Communication-efficient FL}

Multiple collaborative intelligent MUEs can form a network for computing, where the network nodes collaborate to train neural networks by sharing model parameters without exposing private data~\cite{Zhao2023}. The key issue is efficiently training neural network models while fully exploiting the communication performance and computational power among nodes. In the ICSAC architecture, networked sensing provides sensing information repositories. Exploiting the prior information on the positions of MUEs and CSI estimation, distributed MUEs can quickly select cooperative nodes with better communication quality for collaborative training, reducing control overhead and application latency.
Besides, with the sensing results, FL can be used for resource allocation predictions for solving wireless communication problems.

\subsubsection{Sensing-enhanced Semantic Communication}

Neural networks, auto-encoders/decoders, and other related techniques can extract the compressed model information as semantic features, and the goal is to maximize the semantic information while minimizing the data size~\cite{Zhao2023}. In machine-type applications, motion control is a prominent requirement. Suppose the position, velocity, and posture information of the information source and sink MUEs can be integrated into the semantic extraction process. In that case, there is the potential to generate higher-efficiency semantic information for machine motion control.

\textbf{Challenges to the Sensing-empowered ICAC Techniques:} The accuracy, coverage, and timeliness of sensing information significantly impact the above sensing-empowered ICAC techniques. Ensuring timely and comprehensive updates of sensing information is a challenging issue to address. Moreover, the efficient exploitation of the communication network performance for distributed computing is also a great challenge in the highly dynamic environment.

\subsection{ICSAC Techniques}

Finally, we present the crucial techniques for achieving the ICSAC reciprocity mentioned in Section~\ref{sec:Reciprocity_CSC}, based on the computing-empowered ISAC and sensing-empowered ICAC techniques.

\subsubsection{Digital Twin}

Based on the computing-empowered ISAC and sensing-empowered ICAC techniques mentioned in Sections~\ref{sec:ISAC_techniques} and \ref{sec:ICAC_techniques}, MUEs and APs can cooperate to obtain sensing data with high timeliness using networked sensing and preprocess the distributed sensing data with distributed computing, which can be aggregated and updated hierarchically at the edge cloud and core cloud levels, forming a global sensing information repository that contains information for constructing virtual models of physical entities. Therefore, based on the sensing information repository, we can utilize AI techniques embedded in the distributed or central ICAC computing infrastructures to build a digital virtual world corresponding to the physical world, enabling the realization of digital twin technology.

\subsubsection{Metaverse}
The high-fidelity digital world formed by digital twins can provide interfaces for intelligent interaction with humans. After semantic compilation, human instruction can interact with the digital twin world. Humans can collaborate or communicate in the digital world or conduct remote tasks via AI agents in the digital world. AI agents can generate network control signals based on human-machine interaction instructions to remotely operate the IMs through the ICSAC network infrastructure, enabling to change the environment in the physical world. Moreover, based on the huge sensing information repository aggregated at the core cloud, the core cloud can use a generative adversarial network (GAN) or large pre-training model (LPTN) to generate synthetic media information for HUEs to enhance the experience of 6G immersive multimedia applications, such as holographic interactions and embodied intelligence.


\textbf{Challenges to the ICSAC Techniques:} The interaction between the digital world and the physical world requires high timeliness and data reliability to ensure the realism of the digital world and guarantee user experience. Additionally, it is necessary to significantly enhance the security and privacy of network access to ensure the security of human-machine interaction in the real world.

Finally, the features, enabling technologies, and challenges for achieving the aforementioned ISAC, ICAC, and ICSAC reciprocity are summarized in Fig.~\ref{fig:ICSAC_table}. 

\begin{figure}[!t]
	\centering
	\includegraphics[width=0.27\textheight]{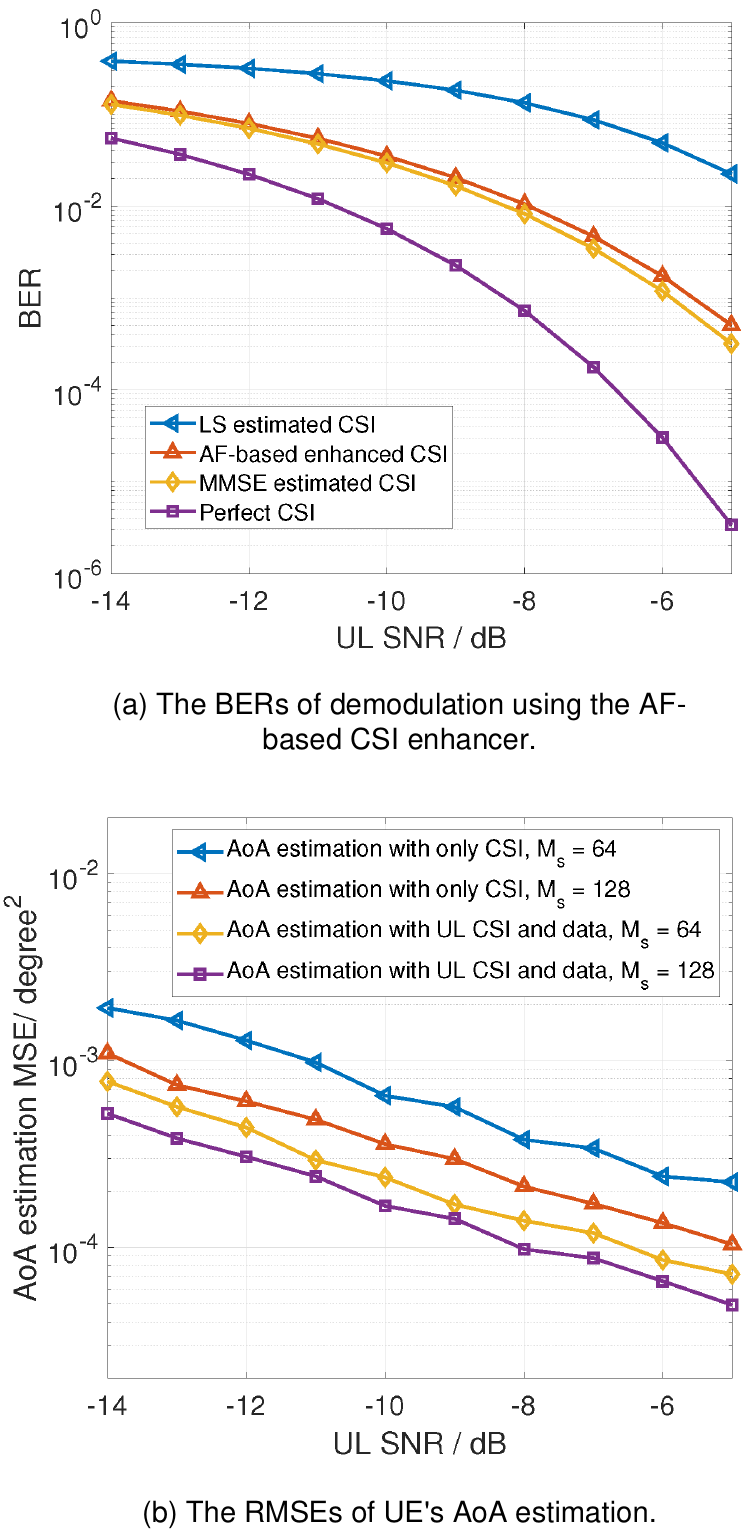}%
	\DeclareGraphicsExtensions.
	\caption{The mutual communication and sensing performance enhancement. (a) The BER performance is enhanced by the AF-based CSI enhancer. (b) The AoA sensing performance is improved by using the combined UL CSI estimation and received data signals that are enhanced by the AF-based CSI enhancer.}
	\label{fig:MSE_ISAC}
\end{figure}


\section{Evaluation Results} \label{sec:evaluation}





In this section, we present the evaluation results of several typical enabling technologies mentioned above to demonstrate the feasibility of the ICSAC framework. 

{\color{black} Fig.~\ref{fig:MSE_ISAC} presents the mutual communication and sensing performance enhancement using the computing-empowered ISAC techniques in Section~\ref{sec:ISAC_techniques}. Fig.~5(a) presents the bit error rate (BER) performance of demodulation using the AF-based CSI enhancer mentioned in Section~\ref{sec:ISAC_techniques}, LS, and MMSE methods when the array size of AP is $8 \times 8$ under 4-QAM modulation. The AF-based CSI enhancer exploits the AoA sensing results as prior information to improve the CSI estimation accuracy. Fig.~5(b) further presents the AoA estimation performance enhancement using the combined CSI estimation and the demodulated data signals enhanced by the AF-based CSI enhancer. Specifically, the improved BER performance enhances the accuracy of data decoding, which further improves the estimation of correlation between the received data signals and the decoded data, thus improving the AoA estimation performance. Fig.~5(a) shows that the AF-based enhancer requires about 3.5 dB lower SNR than the LS method, but 0.2 dB higher SNR than the MMSE method to achieve the given BER. With the enhanced communication reliability, Fig.~5(b) shows that the MSEs of AoA estimation based on the combined CSI and received data signals are lower than those only based on pure CSI estimation. Moreover, the larger number of packets, $M_s$, results in lower AoA estimation MSEs. This is because a larger number of symbols can accumulate more effective energy for sensing.}


\begin{figure}[!t]
	\centering
	\includegraphics[width=0.27\textheight]{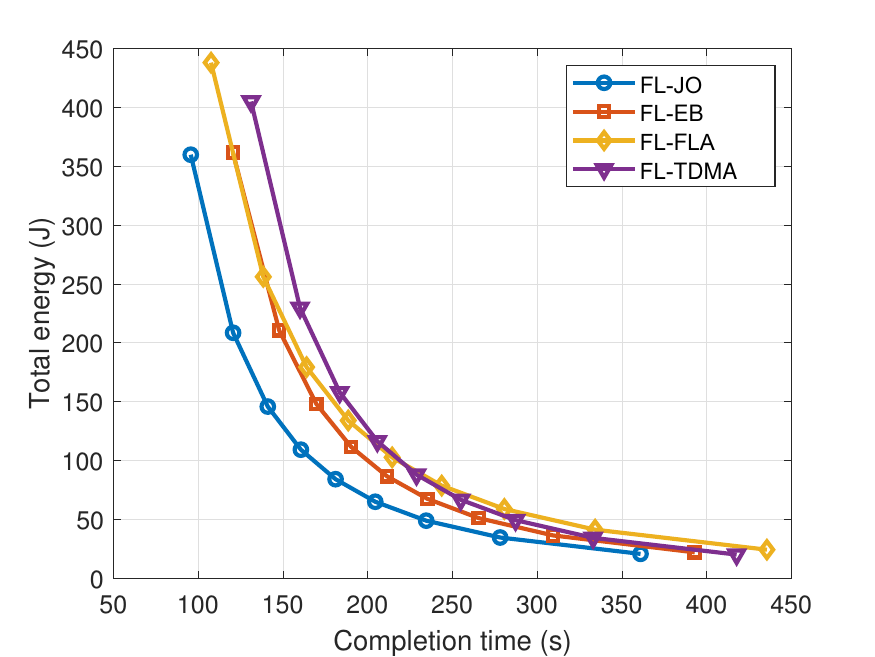}%
	\DeclareGraphicsExtensions.
	\caption{Total energy versus the completion time for various FL resource allocation schemes. `FL-JO' stands for the FL with joint optimization of bandwidth and local accuracy, `FL-EB', `FL-FLA', and `FL-TDMA' stand for the schemes with equal bandwidth allocation for multiple users, with fixed local accuracy of the sensing task, and with users using TDMA, respectively.}
	\label{fig:FL}
\end{figure}

The performance of the communication-efficient FL mentioned in Section~\ref{sec:ICAC_techniques} is shown in Fig. \ref{fig:FL}, where the total energy consumption versus the completion time for various FL resource allocation schemes for sensing is provided. The total energy/time includes the parts for both communication and computation. 
It is shown that the total energy decreases as the total completion time increases.
This is because a long completion time can ensure a small transmit and computation power, thus reducing the total system energy. 
The figure also shows that FL-JO achieves the best performance, which indicates the superiority of joint communication and computation design.

\section{Conclusion} \label{sec:conclusion}
In this article, we have proposed an ICSAC framework for 6G to realize the reciprocity among CSC functions to enhance the communication reliability and latency, timeliness and accuracy of sensing information acquisition, and privacy and security of computing to support the IMN and immersive applications. Furthermore, we have elaborated on the potential enabling technologies and outlined the corresponding challenges to achieve them. Evaluation results show the feasibility of the proposed enabling technologies. This article has drawn a blueprint for innovations of 6G IMN systems. In the future, the efficient algorithms that can exploit the reciprocity of CSC functions to enhance the overall performance and the theoretical bounds for CSC performance are the fundamental problems to be resolved.

\ifCLASSOPTIONcaptionsoff
  \newpage
\fi

\end{document}